\documentclass[runningheads]{llncs}
\usepackage{graphicx}
\usepackage{xcolor}
\usepackage{etex}
\usepackage{hyperref}
\usepackage[utf8]{inputenc}
\usepackage[T1]{fontenc}
\usepackage[ngerman,english]{babel}
\usepackage{wrapfig}
\usepackage{times}
\usepackage{paralist}
\usepackage{graphicx}
\usepackage{caption}
\usepackage{subfig}
\usepackage{booktabs}
\usepackage{paralist}
\usepackage{amssymb, amsfonts}
\usepackage{amsmath}
\usepackage{amsopn}
\usepackage{adjustbox}
\usepackage{url}
\usepackage{multirow}
\usepackage{relsize}
\usepackage{xfrac}
\usepackage{pdflscape}
\usepackage{cite}
\usepackage{tabularx}
\usepackage{mdframed}
\usepackage{marginnote}
\usepackage{algorithm}
\usepackage{algorithmicx}
\usepackage[noend]{algpseudocode}
\usepackage{inconsolata}
\usepackage{nicefrac}
\usepackage{orcidlink}
\usepackage{xspace}
\usepackage{enumitem}

\newcommand{\mypar}[1]{\smallskip\noindent\textbf{#1.}}
\newcommand{\mypartwo}[1]{\noindent\textit{#1.}}

\begin{document}
\title{From OCEL to DOCEL --\\ Datasets and Automated Transformation }
%

\author{Alexandre Goossens$^{\ddag}$\inst{1}\orcidlink{0000-0001-8907-330X},
Adrian Rebmann$^{\ddag}$\inst{2}\orcidlink{0000-0001-7009-4637}
Johannes De Smedt\inst{1} \orcidlink{0000-0003-0389-0275},\\
Jan Vanthienen\inst{1}\orcidlink{0000-0002-3867-7055}  \and
Han van der Aa \inst{2} \orcidlink{0000-0002-4200-4937}}
%
%
\authorrunning{A. Goossens et al.}

\institute{Leuven Institute for Research on Information Systems (LIRIS), KU Leuven \email{\{FirstName\}.\{LastName\}@kuleuven.be}\\ \and Data and Web Science Group, University of Mannheim, Germany\\\email{\{rebmann|han.van.der.aa\}@uni-mannheim.de}}
\maketitle              
\begin{abstract}
Object-centric event data represent processes from the point of view of all the involved object types. This perspective has gained interest in recent years as it supports the analysis of processes that previously could not be adequately captured, due to the lack of a clear case notion as well as an increasing amount of output data that needs to be stored. 
Although publicly available event logs are crucial artifacts for researchers to develop and evaluate novel process mining techniques, the currently available object-centric event logs have limitations in this regard. Specifically, they mainly focus on control-flow and rarely contain objects with attributes that change over time, even though this is not realistic, as the attribute values of objects can be altered during their lifecycle.
This paper addresses this gap by providing two means of establishing object-centric datasets with dynamically evolving attributes.
First, we provide event log generators, which allow researchers to generate customized, artificial logs with dynamic attributes in the recently proposed DOCEL format.
Second, we propose and evaluate
an algorithm to convert OCEL logs into DOCEL logs, which involves the detection of event attributes that capture evolving object information and the creation of dynamic attributes from these. 
Through these contributions, this paper supports 
the advancement of object-centric process analysis by providing researchers with new means to obtain relevant data to use during the development of new techniques.


\keywords{Object-centric processes \and OCEL \and DOCEL \and  Log Generator.}
\end{abstract}

\renewcommand{\thefootnote}{\fnsymbol{footnote}}
\footnotetext[5]{Joint first authors}
\renewcommand{\thefootnote}{\arabic{footnote}}
\addtocounter{footnote}{-2}%
\vspace{-2.5em}
\section{Introduction}
\label{sec:introduction}
Organizations often operate in complex environments with multiple objects interacting and participating in the same business process.
To capture these different perspectives, the concept of object-centric processes has been proposed, in which multiple object types participate over the course of a business process \cite{van2001proclets}.
In recent years, object-centric process mining has gained increasing interest in the research community with the introduction of novel event log formats such as eXtensible Object-Centric (XOC) logs \cite{li2018extracting}, Object-Centric Behavioral Constraint (OCBC) models \cite{van2017object}, and Object-Centric Event Logs (OCEL) \cite{ghahfarokhi2020ocel}. Especially, OCEL is currently the most used object-centric event log format with its own evaluation metrics \cite{adams2021precision}, visualization tool \cite{ghahfarokhi2022python} and various analysis techniques \cite{berti2022filtering,adams2022defining}.

However, OCEL has some limitations by design. Particularly, the relationships between objects and attributes are not strictly defined and attributes that can change in value over time are challenging to deal with in OCEL \cite{goossens2022enhancing}. This makes
the analysis of attribute change difficult, because it is not always clear which event or object manipulated the attribute value, e.g. whether an update event changed the quantity of an order or the price of a product.
To overcome this, there have been efforts to establish a new format for object-centric event data such as Data-aware Object-Centric Event Logs (DOCEL) \cite{goossens2022enhancing}. The main aspect DOCEL introduced is the notion of static and dynamic attributes \cite{goossens2022enhancing}. Static attributes are attributes that do not change over the course of a business process and can either be linked to an event or to an object. Conversely, dynamic attributes are attributes that can change over the course of a business process and are linked to both an object and to an event with the use of foreign keys. As of now, no general consensus has been reached, however, regarding the exact set of entities and relationships a meta model of such a data format shall have. Currently, the Object-Centric Event Data (OCED) is under development and will surely take into account the explicit representation of object evolution and their attributes.\footnote{https://www.tf-pm.org/resources/oced-standard} 

However, despite the benefits of moving to object-centric data with dynamic attributes, no datasets are available that can actually be leveraged by researchers for the development of process mining techniques that account for such attributes. 

This paper addresses this issue in two ways: 

\begin{compactenum}
\item We propose two process-specific log generators to create customized object-centric event logs in the DOCEL format~\cite{goossens2022enhancing}, which also generate dynamic object attributes, i.e., they create objects with attributes that change over time. 
\item We propose an algorithm that takes an existing OCEL event log and automatically transforms it into an event log in the DOCEL format. Combined with earlier work, this algorithm can also be used to transform flat XES~\cite{xes2014ieee} logs into DOCEL logs. 
\end{compactenum}

\noindent
The remainder of this paper is structured as follows: \autoref{sec:motivation} motivates the need for data sets with dynamic object attributes. \autoref{sec:datasets} presents our log generators to create DOCEL logs. \autoref{sec:algorithm} then describes our algorithm to transform OCEL into DOCEL logs. \autoref{sec:evaluation} uses data sets created using our log generators to evaluate the algorithm and shows how even real-life event data captured in XES format can be transformed into DOCEL and thus be used for research on object-centric process mining. Finally, \autoref{sec:related} reflects on related work and \autoref{sec:conclusion} concludes the paper.

\section{Motivation}
\label{sec:motivation}
Over the course of the execution of a business process, the objects involved in its execution may undergo changes, such as creating, updating or deleting, which results in attribute values changes throughout an object's lifecycle.
The eXtensible Event Stream (XES) event log format, with its traditional case-based view of processes, addresses this issue by directly linking attributes to the events that manipulate them \cite{xes2014ieee}. 
Because each event belongs to exactly one case, it is unambiguously clear which attribute was manipulated by a specific event as well as to which trace it belongs.

However, in object-centric processes, events may refer to any number of objects~\cite{adams2022defining}. In OCEL, this is implemented using an event table storing all attributes that are manipulated by events and an objects table storing all objects and their (static) attributes. 
Unfortunately, with OCEL it is not possible to uniquely identify to which object a dynamic attribute belongs, neither through its events table nor its objects table \cite{goossens2022enhancing}. 
To illustrate this, consider the event table of an OCEL log shown in \autoref{table:exampleocellog} and its objects table in \autoref{table:objects}, which cover the (simplified) handling of two orders. First, an order is created, then the ordered items are picked, before they are sent. Between creating an order and sending items, orders can be updated, i.e., items may be added or removed.
Besides the common \texttt{EventID}, \texttt{Activity}, and \texttt{Timestamp} attributes and the references to the instances of different object types associated with each event, the log contains an additional event attribute, \texttt{Value}. While this attribute is associated with events, it is clear that it actually refers to objects. 
Specifically, it refers to the current value of the order the event is associated with. 
This is left implicit by the OCEL format making it impossible for automated process analysis techniques to properly leverage this information. 
For instance, by looking at event $e_6$, it is unclear whether the \emph{Value} refers to the current value of the order or the value of the item that is removed. 
This makes analyzing attribute changes in object-centric business processes difficult using OCEL.

\begin{table}[htbp]
 \vspace{-1em}
    \begin{minipage}{0.6\textwidth} 
        \centering
        \begin{tabular}{cccccr}
        	\toprule
        	\textbf{\texttt{ID}}& \textbf{\texttt{Activity}} & \textbf{\texttt{Timestamp}}&  \textbf{\texttt{Orders}} &  \textbf{\texttt{Items}} &  \textbf{\texttt{Value}}\\
        	\midrule
        	$e_1$&Create order & 05-20 09:07&$\{o_1\}$&$\{i_{1},i_{2}\}$&$\emph{100}$\\	
        	$e_2$&Pick items & 05-23 14:20&$\{o_1\}$&$\{i_{1},i_{2}\}$& $\emph{100}$\\
        	$e_3$&Create order & 06-03 19:17&$\{o_2\}$&$\{i_{3}\}$&$\emph{60}$\\
        	$e_4$&Pick items & 06-04 15:20&$\{o_2\}$&$\{i_{3}\}$&$\emph{60}$\\
        	$e_5$&Update order & 06-04 18:11&$\{o_1\}$&$\{i_{1}\}$&$\emph{70}$\\
        $e_6$&Remove item & 06-05 11:48&$\{o_1\}$&$\{i_{2}\}$&$\emph{70}$\\
        	\bottomrule
        \end{tabular}
        \caption{Events of an OCEL log.}
        \label{table:exampleocellog}
    \end{minipage}
    \begin{minipage}{0.4\textwidth} 
        \centering
        \begin{tabular}{cl}
            \toprule
	       \textbf{\texttt{Type}}& \textbf{\texttt{Instances}}\\
            \midrule
             \vspace{.65em}
            Orders &  $\{o_1,$\\
            &\hphantom{\{}$o_2\}$\\
            \midrule
            Items &$\{i_{1}(\texttt{Weight: 24}),$\\
            &\hphantom{\{}$i_{2}(\texttt{Weight: 99}),$ \\
            &\hphantom{\{}$i_{3}(\texttt{Weight: 10})\}$ \\
            \bottomrule
        \end{tabular}
        \caption{Objects of the log}
        \label{table:objects}
    \end{minipage}
    \vspace{-1.5em}
\end{table}

The DOCEL format addresses this issue with the notion of static and dynamic attributes. Static attributes do not change over the course of a business process, whereas dynamic object attributes can change value over the course of a process and are linked to both an event and an object.
For instance, consider the running example, where orders have a \texttt{Value} attribute. 
If an order changes during its lifecycle, i.e., between its first and last occurrence in the execution of a business process, e.g., because items are added later on, the value attribute changes. 
DOCEL explicitly links this change to the order itself and the event that caused it.
Unfortunately, despite the increased interest in object-centric processes, there are currently no event logs available describing processes with dynamic attributes nor are there tools available to generate such object-centric logs.

\section{DOCEL Dataset Generators}
\label{sec:datasets}
This section introduces the DOCEL log generators for two artificial, yet realistic processes. 
First, we introduce the two processes that are simulated by the log generators. 
These processes contain dynamic attributes as well as various AND and OR-gateways, and loops, which are also found in real-life processes. 
Second, we explain the log generators and their tunable parameters.

\subsection{Process descriptions}
In this section, we describe the processes that serve as a basis for our log generators. The  process models can be found alongside the Python notebooks that implement the generators\footnote{Available at \url{https://github.com/a-rebmann/ocel_to_docel/tree/main/notebooks}}. In this paper, we limit ourselves to a textual description for space reasons.

\subsubsection{Order-to-delivery process}
The order-to-delivery process\footnote{This process is based on the running example OCEL log available on \url{ocel-standard.org}} contains 5 object types and various attributes that we summarize in \autoref{process1objecttypes}.

\begin{table}[!t]
\vspace{-1.5em}
\centering
\caption{Object types order-to-delivery process}
\label{process1objecttypes}
\begin{tabular}{lll}
\hline
\textbf{Object Type}  & \textbf{Static Attributes}  & \textbf{Dynamic Attributes} \\ \hline
\textbf{Customer}     & Name, Bank Account          & Customer Address             \\
\textbf{Order}        &        /                     & Weight, Order Price         \\
\textbf{Product Type} & Product Name, Price, Weight &   /                          \\
\textbf{Item}         & Price, Weight               &    /                         \\
\textbf{Packages}     & Price, Weight               &  /  \\ \hline                        
\end{tabular}
\vspace{-1em}
\end{table}

\textit{The process starts with the customer adding items to an order which increases the value and the weight of an order. Each item is of a certain product type and inherits its weight and price. Once the order is placed, the items are picked. Before paying the order, the customer is still allowed to remove items from an order. If that is the case, the items are removed and the value and weight of an order are updated. Once the order is paid, the package is created with a weight and price and sent out. However, a customer might change their delivery address in which case the delivery fails and has to be re-executed. Moreover, every delivery has a very small chance of failing due to unforeseen circumstances. Once the package is successfully delivered, the process ends.}

\subsubsection{Shipping-method process}
The shipping-method process covers orders that have different shipping modes depending on various factors. It is based on the process described in \cite{desmedt2017towards} and contains three object types whose attributes are summarized in Table \ref{OTshippingprocess}.

\begin{table}[!t]
\centering
\caption{Object types shipping-method process}
\label{OTshippingprocess}
\begin{tabular}{lll}
\hline
\textbf{Object Type}  & \textbf{Static Attributes} & \textbf{Dynamic Attributes}    \\ \hline
\textbf{Customers}    & Name, Bank Account         & /                              \\
\textbf{Product Type} & Value, Fragile             & /                              \\
\textbf{Orders}       & Quantity                   & Value, Refund, Shipping Method \\ \hline
\end{tabular}
\vspace{-1.5em}
\end{table}

\textit{A customer places an order with a quantity for each product type. This is then received by the company which has to manually confirm the purchase at which point the value of the order is also determined. Once the purchase is confirmed, the products are retrieved from the warehouse and added to the package. In case, the product type is fragile (indicated with a binary value), the product is first wrapped with some protection. Next, the customer has to confirm the shipping information, after which the shipping method is determined. If the package contains a fragile product or its value exceeds a certain amount, the package is shipped with a courier. 
Otherwise the package is shipped by mail. Simultaneously, an invoice is sent to the customer. Once the package has arrived, the customer determines their satisfaction. If the customer is satisfied the process is ended after the order has been filed.
In case, the customer is dissatisfied, the customer requests a refund, setting the binary refund value to 1. Next, the customer confirms the shipping information, which will be handled using an express courier.The company sends a recollect letter after which the customer returns the package. Once the package is returned, the company refunds the customer. Once the package has arrived, the customer determines their satisfaction (assumed to be positive here) after which the order is filed and the process ends.}

\subsection{Log Generators}

\begin{table}[b!]
    \vspace{-1em}
    \caption{Summary of functionality in the order-to-delivery Log Generator}
    \label{tab:log_generator_ordertodelivery}
    \begin{adjustbox}{width=\textwidth}
    \centering
    \begin{tabular}{|l|p{10cm}|}
        \hline
        \textbf{Tunable parameter} & \textbf{Description of Functionality} \\
        \hline
        Customer Addresses and Names & Randomly generated for each log entry along with randomly generated bank account details. \\
        \hline
        Products & Taken from a fixed list directly obtained from the OCEL log of X. \\
        \hline
        Start Timeframe & Allows defining the start time for the process. \\
        \hline
        Time Between Events & Allows adjusting the time interval between consecutive events. \\
        \hline
        Number of Orders & Can be changed to generate logs with varying numbers of orders. \\
        \hline
        Max Number of Products & Can be adjusted to set a maximum limit for the number of products in an order. \\
        \hline
        Max Number of Items & Can be modified to set a maximum limit for the total number of items in an order. \\
        \hline
        Probability of Removing an Item & Can be adjusted to control the likelihood of an item being removed from an order. The higher the value, the more likely an item is removed. \\
        \hline
        Probability of Changing an Address & Allows changing the probability of an address being modified in the log entries. The higher the probability, the more likely the address will be changed. \\
        \hline
        Probability of Failing a Delivery & Can be altered to control the likelihood of a delivery failure occurrence. The higher the probability, the more likely the probability will be changed. \\
        \hline
        
        \hline
    \end{tabular}
\end{adjustbox}
    \vspace{-1em}
\end{table}

\begin{table}[t!]
    \vspace{-1em}
    \caption{Summary of functionality in the shipping method log generator}
    \label{tab:log_generator_shippingmethod}
    \centering
    \begin{adjustbox}{width=\textwidth}
    \begin{tabular}{|l|p{10cm}|}
        \hline
        \textbf{Tunable parameters} & \textbf{Description of Functionality} \\
        \hline
        Number of Products & Allows adding new products with custom values for the price and the possibility of fragility. \\
        \hline
        Number of Customers & Enables creating new customers with randomly generated names and bank accounts. \\
        \hline
        Lists of People Executing Activities & Allows adapting the lists of people who execute the company's activities. \\
        \hline
        Start Timeframe & Allows defining the start time for the process. \\
        \hline
        Time interval Between Events & Allows adjusting the time interval between consecutive events. \\
        \hline
        Order Value Threshold & Can be changed to determine the shipping method based on the order's value. \\
        \hline
        Number of Orders & Can be changed to generate logs with varying numbers of orders. \\
        \hline
        Probability of Refund & Can be adapted to determine the likelihood of a refund, indicating customer satisfaction with the purchase. The higher the value the more satisfied the customer is. \\
        \hline
    \end{tabular}
\end{adjustbox}
    \vspace{-1em}
\end{table}

To allow the research community to generate as many different logs as desired for different research needs, 
we propose log generators that allow users to change various parameters that influence the generated logs, as summarized in Tables \ref{tab:log_generator_ordertodelivery} and \ref{tab:log_generator_shippingmethod}.

In both log generators, it is possible to change the amount of object instances of all the object types involved in the process as well as changing the time between events and initial time frame of a process. Beyond that, it is possible to change various probabilities of events happening in the process such as removing an item or asking for a refund. 
Both log generators can generate DOCEL logs in a spreadsheet format, which is intuitive to understand because every table (events, objects and dynamic attribute tables) can be stored in a separate sheet. Note that, because we provide the generators as Python notebooks, it is possible to change the logs beyond the options described in the tables with some minimal Python coding.

Finally, to be able to evaluate our automated OCEL-to-DOCEL transformation algorithm (cf .\autoref{sec:algorithm}), we also included the option to create an OCEL log for a generated DOCEL log.
If OCEL logs are created, the dynamic attributes are directly linked to the events, therefore, losing the clear information on object-attribute allocations.

\section{Transforming OCEL to DOCEL}
\label{sec:algorithm}
This section presents our proposed algorithm to transform an OCEL event log into a DOCEL formatted one, achieved by detecting  dynamic object attributes and assigning them to the appropriate object instances.
Our algorithm takes as  input an OCEL formatted event log $L$, which comprises a set of events recorded by an information system.
Each event $e \in L$ is a tuple $e = (eid, act, ts, OI, AV)$, with $eid$ the event's id, $act$ its activity, $ts$ its timestamp, $OI$ a set of object instances, and $AI$ a set of attribute-value pairs. 
Each object instance $oi \in OI$ is a tuple ($oid$, $type$), where $oid$ is the instance's identifier and $type$ its type, whereas each attribute-value pair $(a,v) \in AI$ relates an attribute value $v$ to an attribute name $a$.
We denote the set of object types that occur in $L$ as $\mathcal{O}_T^L$ and the set of event attribute names as $\mathcal{A}_T^L$.

\begin{figure}[h!]
\vspace{-1em}
    \centering
    \includegraphics[width=\linewidth]{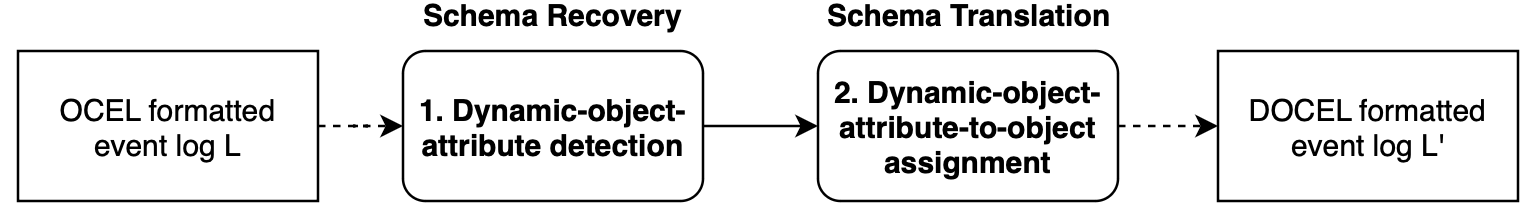}
    \caption{Algorithm overview.}
    \label{fig:approach}
        \vspace{-2em}
\end{figure}

As visualized in \autoref{fig:approach}, our algorithm applies two steps to transform an OCEL log $L$ into a DOCEL formatted log $L'$.
Step 1, \emph{Dynamic object attribute detection}, aims to detect dynamic object attributes in the event attributes of the input log and match them with the object type they refer to. 
Based on that matching, Step 2, \emph{Dynamic-object-attribute-to-object assignment}, creates object attributes and associates these with the individual object instances they relate to, establishing a DOCEL log that makes these relationships explicit. 
In the remainder of this section, we describe these steps in detail.

\subsection{Dynamic object attribute detection}
\label{sec:alg:step1}

This step aims to determine whether an attribute $a \in \mathcal{A}_T^L$ is a dynamic object attribute of an object type $t \in \mathcal{O}_T^L$, resulting in a set of attribute-object-type matches $M$.

For this step, recall that dynamic object attributes are attached to events in the OCEL format, even though they capture information about an object associated with an event rather than relate to the event itself.
For instance, although an \emph{Update order} event has a \texttt{Value} attribute in \autoref{table:exampleocellog}, this attribute actually captures the (new) value of the order, not of the event.

\mypar{Identifying candidate object types}
For each attribute $a \in \mathcal{A}_T^L$, our algorithm first identifies a set of candidate object types $O_a \subseteq \mathcal{O}_T^L$. To determine if an  object type $t \in \mathcal{O}_T^L$ is a candidate type for attribute $a$, our algorithm checks if it meets two requirements:
\begin{compactitem}
    \item \textit{Object-attribute co-occurrence:} First, our algorithm checks if every occurrence of attribute $a$ can be related to exactly one instance of object type $t$. This means that $L$ cannot contain an event $e$ for which $(a, v) \in e.AV$ and for which $e.OI$ either does not contain any object instance of type $t$ or contains multiple of them. For instance, for the running example type \textit{item} is not a candidate for attribute \texttt{Value}, since events $e_1$ and $e_2$ each refer to two items, but only contain a single \texttt{Value}.
    \item \textit{Observed attribute changes:} Second, our algorithm checks if attribute $a$ is actually dynamic (with respect to type $t$), i.e., if its value is observed to change during an object's lifecycle. For this, our algorithm checks if $L$ contains (at least) two events, $e$ and $e'$, which both contain the same object instance of type $t$, but are associated with different values for attribute $a$.
\end{compactitem}
\noindent Any object type $t$ that passes both checks is added to the set of candidate object types $O_a$.
If, after checking all object types in $\mathcal{O}_T^L$, $O_a$ contains exactly one candidate type $t$, the match $(a, t)$ is added to $M$. Instead, if $O_a$ does not contain any candidates, then $a$ is not a dynamic attribute, whereas, if $O_a$ contains more than one candidate, our algorithm turns to the disambiguation procedure described next.

\mypar{Disambiguating candidate types}
An attribute $a$ can have multiple candidate object types in $O_a$ if certain object types always occur together for events. For example, if every event in a log $L$ is associated with both a \textit{customer} and an \textit{order}, an attribute such as \texttt{Value} would have both of these types as a candidate, since each occurrence of \texttt{Value} can be associated with exactly one \textit{customer} and one \textit{order}.

To be able to match $a$ to a single object type $t \in O_A$ in such cases, our algorithm employs two disambiguation strategies:
\begin{compactitem}
    \item \textit{Relation-based selection:} Our algorithm first checks for 1:N relationships between object types in $O_a$, aiming to assign $a$ to a more fine-granular candidate object type.

    For illustration, consider a \texttt{Refund} attribute and two object types, \emph{customer} and \emph{order} in $O_{\texttt{Refund}}$. All events with a value for \texttt{Refund} refer to one customer and one order. Matching \texttt{Refund} to \emph{customer} would obfuscate the relation between single orders and their \texttt{Refund} if a customer places multiple orders.
    Therefore, the algorithm tries to identify cases for each $t \in O_a$, where for two events $e_1, e_2 \in L$ an instance $oi_t$ of $t$ occurs together with two different instances $oi_{t'_1}$ and $oi_{t'_2}$ of another type $t' \in O_a$ and, if so, removes $t$ from the candidates. If a single candidate $t$ remains, the match $(a, t)$ is added to $M$.

    \item \textit{Name-based selection:} If, after the relation-based selection, $O_a$ still contains more than one candidate, our algorithm finds the most similar object type among the candidates based on $a$'s name.
    For this, our algorithm quantifies the semantic similarity $\texttt{sim}(a, t) \in [0,1]$ between $a$ and each type $t \in O_a$ by computing the cosine similarity between sentence embeddings, obtained from a pretrained \emph{Sentence Transformer}~\cite{reimers2019sentencebert}.
    These embeddings are specifically designed to capture semantically meaningful representations on the level of (short) sentences rather than individual words, making them highly suitable for our purpose.
    
    Given these similarity scores, our algorithm determines if there is a type $t \in O_a$ for which the similarity score is distinctly higher (according to a threshold $\tau$, set to 0.1 by default) than the scores of the other types, i.e., if $\texttt{sim}(a, t) > \texttt{sim}(a, t') + \tau$ for each $t' \in O_a \setminus \{t\}$.
    If such a distinctively similar type $t$ is found, the match $(a, t)$ is added to $M$, otherwise $a$ is not matched to any object type.
\end{compactitem}

\noindent
The set of attribute-object-type matches $M$ then serves as input to the next step.

\subsection{Dynamic-object-attribute-to-object assignment}
Based on each match $(a,t) \in M$, Step 2 creates an object attribute $a_{(e,oi)}=(vid, eid,$ $oid, a)$ for each event $e$ that changes $oi$'s value for $a$. 
The algorithm first assign $a_{(e,oi)}$ a unique identifier $a_{(e,oi)}.vid$. Then it sets its attribute-value pair,  $a_{(e,oi)}.a=(a, v)$ with $(a, v) \in e.AI$, its event identifier, $a_{(e,oi)}.eid=e.eid$, and its object instance identifier, $a_{(e,oi)}.oid$ $=oi.id$ with $oi \in e.OI \land oi.type=t$.
For the attribute \texttt{Value} of our running example, this creates the object attribute table visualized in \autoref{tab:attributetbale}.

\begin{wraptable}{r}{5cm}
\vspace{-2em}
\centering	
\smaller
\caption{\texttt{Value} attribute table.}
\begin{tabular}{cccr}
	\toprule
	\textbf{\texttt{ValueID}}& \textbf{\texttt{OrderID}}& \textbf{\texttt{EventID}} & \textbf{\texttt{Value}}\\
	\midrule
	v1 & o1& e1& 100 \\
    v2 & o2& e3& 60\\
	v3 & o1& e5& 70\\
	\bottomrule
\end{tabular}
\label{tab:attributetbale}
\vspace{-2.5em}
\end{wraptable}

Finally, the algorithm removes the attributes for which a match was found from the events' set of attribute-value pairs and returns the established DOCEL log. For the example, it would, therefore,  remove the \texttt{Value} attribute from the events in \autoref{table:exampleocellog} and return the resulting event table, the unchanged objects table, and the newly created object attribute table (\autoref{tab:attributetbale}).

\section{Evaluation}
\label{sec:evaluation}
We implemented our algorithm in Python and performed evaluation experiments to assess our algorithm's capability to accurately transform OCEL logs into DOCEL logs (\autoref{sec:experiments}). 
Afterwards, we show how it can be used in combination with an existing approach to also transform XES event logs into DOCEL logs (\autoref{sec:xesdocel}).
The implementation, evaluation data, and generated DOCEL logs are available in our repository\footnote{\url{https://github.com/a-rebmann/ocel_to_docel}}.

\subsection{Experiments}
\label{sec:experiments}
We assess whether our algorithm is able to correctly detect the dynamic attributes in an OCEL log and transform it into a DOCEL log.

\mypar{Datasets}
We use OCEL and DOCEL event logs generated using our log generators  (cf. \autoref{sec:datasets}) by simulating the process execution for 100 orders per scenario. The OCEL logs serve as input to our algorithm, whereas the DOCEL logs serve as a gold standard, i.e., the correct transformation result.
The  OCEL logs obtained in this manner differ in their number of events, objects, object types, and attributes as shown in \autoref{table:evaldata}. 

\begin{table}[!h]
\vspace{-2em}
    \centering
    \caption{Characteristics of the OCEL logs used for the evaluation}
    \label{table:evaldata}
    \small
    \begin{tabular}{lclcc}
        \toprule
        \textbf{ID}  & \textbf{\# Events} &  \textbf{\hphantom{space}Objects} &  \textbf{\# Event att.} & \textbf{\# Dyn. att.}\\
        \midrule
        \textbf{Order to} & 6,014 & Customer (44), Order (100),  Item (3,559), & 7 &  3  \\
        \textbf{Delivery}& &Packages (100), Product Type (20) & &\\
        \midrule
        \textbf{Shipping} & 2,036 & Customer (50), Product Type (3), &  5  &   3 \\
         \textbf{Method}& &  Order (100) & & \\
        \bottomrule
    \end{tabular}
    \vspace{-1em}
\end{table}

\mypar{Setup}
To assess the ability of our algorithm to correctly detect dynamic object attributes in the OCEL event logs, we conduct experiments using two settings:

\noindent
(1) Original attribute names. In this setting, we use all information from the event log as input to our approach.

\noindent
(2) Hidden attribute names. To assess the robustness of our algorithm, we reduce the available information by hiding event attribute names in the OCEL logs. This allows us to assess the dependency of our algorithm on its name-based check (\autoref{sec:alg:step1}).

We measure the performance in terms of precision, recall, and F$_1$-score with respect to the dynamic object attributes in the original DOCEL logs. Using $tp$ to denote the dynamic attributes correctly matched to an object type $fp$ for the dynamic attributes incorrectly matched to an object type, and $fn$ for the dynamic attributes that were wrongly not matched to an object type, we then quantify the precision as $tp/(tp+fp)$, the recall as $tp/(tp+fn)$, and F$_1$-score as the harmonic mean of precision and recall.

\mypar{Results}
We first report on the results of attribute-to-object-type matching, which is the most challenging part, before we report on the attribute value to object assignment.

\mypartwo{Attribute-to-object-type matching}
\autoref{table:ocelresults} reports on the results of the attribute-to-object-type matching per event log.
\begin{table}[h!]
	\vspace{-2em}
	\small
	\centering
	\caption{Results of the attribute-to-object-type matching per OCEL log.}
	\label{table:ocelresults}
	\begin{tabular}
		{l  rccc  rccc}
		\toprule
		\textbf{Log}	& \multicolumn{4}{c}{\textbf{Original Attribute Names}} & \multicolumn{4}{c}{\textbf{Hidden Attribute  Names}}\\
		&
		\multicolumn{1}{c}{\textbf{Count}}&
		\multicolumn{1}{c}{\textbf{Precision}}&
		\multicolumn{1}{c}{\textbf{Recall}}&
		\multicolumn{1}{c}{\textbf{F$_{1}$}}&
		
		\multicolumn{1}{c}{\textbf{Count}}&
		\multicolumn{1}{c}{\textbf{Precision}}&
		\multicolumn{1}{c}{\textbf{Recall}}&
		\multicolumn{1}{c}{\textbf{F$_{1}$}}
		\\
		\midrule
		\textbf{Order to delivery} & 3 & 1.00& 1.00 & 1.00 & 3 &1.00 &0.75 & 0.86\\
		\textbf{Shipping method} & 4 & 0.80 & 1.00 & 0.89& 4 & 0.80 &1.00 &0.89 \\
        \midrule
        \textbf{Average} & 3.5 & 0.90 & 1.00 & 0.95& 3.5 &0.90 &0.88 & 0.88\\
		\bottomrule
	\end{tabular}
	\vspace{-1.5em}
\end{table}

We find that our approach achieves a perfect recall and good precision (0.9) in detecting dynamic object attributes and associating these with the correct object type, when all original information from the input log is available. An in-depth look shows that the only error made is the assignment of the \emph{Resource} attribute of shipping-method process to the \emph{order} type. While unconventional, this assignment is not necessarily problematic, since indeed each time a resource executes a process step, that step relates to a specific order.
Such incorrect assignments could be easily avoided by specifying a set of names reserved for event attributes, e.g., \emph{org:resource} or \emph{org:role} as done in XES logs \cite{xes2014ieee}.
The importance of the duplicate-resolution strategy based on object lifecycles becomes clear for the shipping-method process, where \emph{customer} and \emph{order} are in 1:N relation. Without it, e.g., the \texttt{Refund} attribute would be matched to both \emph{customer} and \emph{order}. This strategy resolves this, correctly matching \texttt{Refund} only to the \emph{order}. Note that the name-based matching strategy could not resolve this, because \emph{refund} is semantically similar to both \emph{customer} and \emph{order}.

When hiding the original attribute names, we find that for shipping-method process the performance remains the same, whereas for order to delivery it drops achieving an F$_1$-score of 0.86 compared to 1.00 with original attribute names. The reason for this is the missing match of the \texttt{Customer Address} attribute to the \emph{customer} object type, because the candidates \emph{order} and \emph{customer} could not be resolved solely using the relation-based disambiguation in this case. 

\mypartwo{Object-attribute-to-object assignment}
We also report on the results of Step 2, i.e., the assignment of object-attributes to objects, for the setting with original attribute names. We provide results for (1) when propagating false positives from Step~1 and (2) when only including object attributes that were correctly matched to object types in \autoref{table:ocelresults2}.

\begin{table}[h!]
	\vspace{-2em}
	\small
	\centering
	\caption{Results of the dynamic-object-attribute-to-object assignment per OCEL log for the setting with original attribute names, with and without propagation of false positives (\emph{fp}s) from Step~1.}
	\label{table:ocelresults2}
	\begin{tabular}
		{l  rccc  rccc}
		\toprule
		\textbf{Log}	& \multicolumn{4}{c}{\textbf{When propagating \emph{fp}s}} & \multicolumn{4}{c}{\textbf{Without propagating \emph{fp}s}}\\
		&
		\multicolumn{1}{c}{\textbf{Count}}&
		\multicolumn{1}{c}{\textbf{Precision}}&
		\multicolumn{1}{c}{\textbf{Recall}}&
		\multicolumn{1}{c}{\textbf{F$_{1}$}}&

    \multicolumn{1}{c}{\textbf{Count}}&
		\multicolumn{1}{c}{\textbf{Precision}}&
		\multicolumn{1}{c}{\textbf{Recall}}&
		\multicolumn{1}{c}{\textbf{F$_{1}$}}
		\\
		\midrule
		\textbf{Order to delivery} & 3,710 & 1.00& 1.00 & 1.00& 3,710 & 1.00& 1.00 & 1.00 \\
		\textbf{Shipping method} &1,886& 0.23 & 1.00 & 0.37&  442& 1.00 & 1.00 & 1.00\\
        \midrule
        \textbf{Average} &2,798  & 0.62 & 1.00 & 0.69 &1,875  & 1.00 & 1.00 & 1.00\\
		\bottomrule
	\end{tabular}
	\vspace{-1em}
\end{table}

We find that when propagating false positives (\emph{fp}s) from Step~1, we achieve an average $F_1$-score of 0.69, a precision of 0.37, and a perfect recall (1.00). 
The lower precision is caused only by the assignment of the \texttt{Resource} attribute to the \emph{order} object type for the shipping-method event log. Given that there are many more resource handovers involved than value changes to the actual object attributes this has a considerable impact on the overall performance scores.  
Apart from this, we achieve perfect scores, though. This can also be seen when disregarding false positives from Step~1. In this case, we achieve perfect scores for the assignment of object attributes to objects.

\subsection{Application: From XES to DOCEL}
\label{sec:xesdocel}
 Recently an approach was proposed to uncover object-centric data in a flat event log, e.g., in XES format, to automatically transform it into a log OCEL format~\cite{rebmannuncovering}.
 Given the limitations of OCEL, we aim to combine our algorithm with this approach to investigate whether XES logs can be transformed into DOCEL.
 
 To this end, we use the BPI Challenge 2017 log~\cite{bpi17}, which captures the loan application process at a financial institute and involves two main types of objects: applications and offers. For a single application, multiple offers can be made, where at some point an offer may be accepted. To make sure we have a dynamic object attribute in this log, we add an \texttt{OfferAccepted} attribute to each event, which is set to \texttt{false} when a new application is created and changes to \texttt{true}, when an offer associated with the application is accepted. The goal is to check if our algorithm can identify this attribute correctly as a dynamic object attribute associated with the \emph{application} object type creating a DOCEL log from the output of the XES-to-OCEL approach.

 When applying our algorithm to the transformed XES log, we find that it indeed correctly matched the \texttt{OfferAccepted} attribute to the application. Based on that it removed the corresponding event attributes and created correct object attributes linked to both the correct application and the event that writes the value.
 Beyond detecting the derived attribute correctly, the algorithm also detected the \texttt{EventOrigin} and the \texttt{Action} attributes as dynamic object attributes of the \emph{application} object type. After inspecting the attribute values, the latter of these matches makes sense, because \texttt{Action} captures the status of the application, which changes throughout its lifecycle. \texttt{EventOrigin}, as its name indicates, captures the origin of the event and was, therefore, falsely associated with \emph{application}.
Nevertheless, this outcome shows the potential of our algorithm to help generate a broader variety of object-centric event logs that consider evolving objects based on available event data. 
The DOCEL log our algorithm created can be found in our repository linked on \autopageref{sec:evaluation}.

\section{Related Work}
\label{sec:related}
The first proposed object-centric event log formats are XOC logs \cite{li2018extracting} together with OCBC models \cite{van2017object}. 
These proposals have scalability issues because of the duplication of attributes and object relations with each executed event. 
Next, OCEL logs were introduced which do not have such scalability issues and are currently the most used object-centric event log format \cite{ghahfarokhi2020ocel} with a lot of dedicated research and tools such as a visualization tool \cite{ghahfarokhi2022python}, fitness and precision metrics \cite{adams2021precision}, clustering analysis \cite{ghahfarokhi2022clustering}, or predictive object-centric process analysis \cite{galanti2022object}.

The conversion of XES logs to OCEL logs has been investigated in \cite{rebmannuncovering}. Next to that a generic approach to extract OCEL logs from SAP systems and relational databases has also been researched in respectively \cite{berti2023generic} and \cite{berti2023genericrelational}. Finally, the extraction of OCEL logs from virtual knowledge graphs was researched in \cite{xiong2022extraction}.

\section{Conclusion}
\label{sec:conclusion}
This paper offers two problem-specific log generators and a transformation algorithm to the research community. The two paramaterizable log generators support both the widely used OCEL format and the more recent DOCEL format, which allows for dynamic attributes and consistent object-attribute allocation. 
To further support the creation of DOCEL logs, 
we also proposed 
an algorithm to convert OCEL logs to DOCEL logs.
This algorithm not only identifies the presence of dynamic attributes, which are better represented in the DOCEL format, but also links the attributes to the correct object types. Our evaluation shows that the algorithm can accurately transform OCEL into DOCEL logs and that, in combination with previous work, it can even be used to convert XES logs to DOCEL logs. 
However, it cannot deal with all situations. For instance, if multiple instances of the same object type are changed by one event (batching), this cannot be handled if these instance have not previously been changed individually.

In the future, we aim to address this limitation. We also plan to develop a comprehensive user interface tool to enhance the user experience. Furthermore, we aim to transform additional XES logs that would benefit from a conversion to DOCEL logs.

%
%
\bibliographystyle{splncs04}
\bibliography{main}

\begin{thebibliography}{10}
\providecommand{\url}[1]{\texttt{#1}}
\providecommand{\urlprefix}{URL }
\providecommand{\doi}[1]{https://doi.org/#1}

\bibitem{van2001proclets}
van~der Aalst, W.M., Barthelmess, P., Ellis, C.A., Wainer, J.: Proclets: A
  framework for lightweight interacting workflow processes. International
  Journal of Cooperative Information Systems  \textbf{10}(04),  443--481 (2001)

\bibitem{van2017object}
van~der Aalst, W.M., Li, G., Montali, M.: Object-centric behavioral
  constraints. arXiv preprint arXiv:1703.05740  (2017)

\bibitem{adams2021precision}
Adams, J.N., van~der Aalst, W.: Precision and fitness in object-centric process
  mining. In: 2021 3rd International Conference on Process Mining (ICPM). pp.
  128--135. IEEE (2021)

\bibitem{adams2022defining}
Adams, J.N., Schuster, D., Schmitz, S., Schuh, G., van~der Aalst, W.M.:
  Defining cases and variants for object-centric event data. arXiv preprint
  arXiv:2208.03235  (2022)

\bibitem{berti2022filtering}
Berti, A.: Filtering and sampling object-centric event logs. arXiv preprint
  arXiv:2205.01428  (2022)

\bibitem{berti2023genericrelational}
Berti, A., Park, G., Rafiei, M., van~der Aalst, W.: A generic approach to
  extract object-centric event data from relational databases  (2023)

\bibitem{berti2023generic}
Berti, A., Park, G., Rafiei, M., van~der Aalst, W.M.: A generic approach to
  extract object-centric event data from databases supporting {SAP} {ERP}.
  Journal of Intelligent Information Systems pp. 1--23 (2023)

\bibitem{desmedt2017towards}
{De Smedt}, J., Hasi\'c, F., Vanthienen, J.: Towards a holistic discovery of
  decisions in process-aware information systems. In: Business Process
  Management. LNBIP, Springer (2017)

\bibitem{bpi17}
van Dongen, B.: {BPI} {C}hallenge (2017).
  \doi{10.4121/uuid:5f3067df-f10b-45da-b98b-86ae4c7a310b}

\bibitem{galanti2022object}
Galanti, R., de~Leoni, M., Navarin, N., Marazzi, A.: Object-centric process
  predictive analytics. arXiv preprint arXiv:2203.02801  (2022)

\bibitem{ghahfarokhi2022python}
Ghahfarokhi, A.F., van~der Aalst, W.: A python tool for object-centric process
  mining comparison. arXiv preprint arXiv:2202.05709  (2022)

\bibitem{ghahfarokhi2022clustering}
Ghahfarokhi, A.F., Akoochekian, F., Zandkarimi, F., van~der Aalst, W.M.:
  Clustering object-centric event logs. arXiv preprint arXiv:2207.12764  (2022)

\bibitem{ghahfarokhi2020ocel}
Ghahfarokhi, A.F., Park, G., Berti, A., van~der Aalst, W.M.: {OCEL}: A standard
  for object-centric event logs. In: ADBIS. pp. 169--175. Springer (2021)

\bibitem{goossens2022enhancing}
Goossens, A., De~Smedt, J., Vanthienen, J., van~der Aalst, W.: Enhancing
  data-awareness of object-centric event logs. In: ICPM Workshops (2022)

\bibitem{xes2014ieee}
Günther, C.W., Verbeek, H.M.W.: {XES} standard definition. IEEE Std  (2014)

\bibitem{li2018extracting}
Li, G., Murillas, E.G.L.d., Carvalho, R.M.d., van~der Aalst, W.: Extracting
  object-centric event logs to support process mining on databases. In: CAiSE.
  pp. 182--199. Springer (2018)

\bibitem{rebmannuncovering}
Rebmann, A., Rehse, J.R., van~der Aa, H.: Uncovering object-centric data in
  classical event logs for the automated transformation from {XES} to {OCEL}.
  In: BPM. pp. 11--16 (2022)

\bibitem{reimers2019sentencebert}
Reimers, N., Gurevych, I.: Sentence-bert: Sentence embeddings using siamese
  bert-networks. In: EMNLP. ACL (11 2019)

\bibitem{xiong2022extraction}
Xiong, J., Xiao, G., Kalayci, T.E., Montali, M., Gu, Z., Calvanese, D.:
  Extraction of object-centric event logs through virtual knowledge graphs. In:
  35th International Workshop on Description Logics, DL 2022, Haifa, Israel,
  August 7-10, 2022 (2022)

\end{thebibliography}

\end{document}